\documentclass[pra,twocolumn,amssymb,superscriptaddress]{revtex4}
\usepackage{amsmath, verbatim, graphicx,color,amsthm}
\usepackage{epsfig}
\usepackage{bbm}
\usepackage[latin1]{inputenc} 
\usepackage{color}

\newcommand{\bra}[1]{\langle #1|}
\newcommand{\ket}[1]{|#1\rangle}
\newcommand{\ketbra}[1]{| #1\rangle \langle #1|}
\newcommand{\ketbrapro}[1]{| #1\rangle \langle \cdot |}

\newcommand{\be}{\begin{equation}}
\newcommand{\ee}{\end{equation}}
\newcommand{\eea}{\end{eqnarray}}
\newcommand{\bea}{\begin{eqnarray}}

\newcommand{\eins}{\openone}
\newcommand{\WW}{\ensuremath{\mathcal{W}}}
\newcommand{\NN}{\ensuremath{\mathcal{N}}}

\newcommand{\TT}{\ensuremath{\mathcal{T}}}

\newcommand{\kommentar}[1]{}

\renewcommand{\vr}{\ensuremath{\varrho}}

\newcommand{\forget}[1]{}

\begin{document}
\title{
Multiparticle entanglement in graph-diagonal states: \\
Necessary and sufficient conditions for four qubits
}

\author{Otfried G{\"u}hne}
\affiliation{Naturwissenschaftlich-Technische Fakult\"at, Universit\"at Siegen,
Walter-Flex-Stra{\ss}e 3, 57068 Siegen, Germany}
\author{Bastian Jungnitsch}
\affiliation{Institut f\"{u}r Quantenoptik und Quanteninformation,
\"{O}sterreichische Akademie der Wissenschaften, Technikerstra{\ss}e
21A, 6020 Innsbruck, Austria}
\author{Tobias Moroder}
\affiliation{Institut f\"{u}r Quantenoptik und Quanteninformation,
\"{O}sterreichische Akademie der Wissenschaften, Technikerstra{\ss}e
21A, 6020 Innsbruck, Austria}
\author{Yaakov S. Weinstein}
\affiliation{Quantum Information Science Group, MITRE, 260 Industrial Way West,
Eatontown, NJ 07724 USA}

\date{\today}
\begin{abstract}
The characterization of genuine multiparticle entanglement is important
for entanglement theory as well as experimental studies related to quantum
information theory. 
Here, we completely characterize genuine multiparticle entanglement for
four-qubit states diagonal 
in the cluster-state basis. In addition, we give a complete characterization of
multiparticle entanglement
for all five-qubit graph states mixed with white noise, for states diagonal 
in the basis corresponding to the five-qubit Y-shaped graph, and for a family of
graph states
with an arbitrary number of qubits.
\end{abstract}

\maketitle

\section{Introduction}
The characterization of multiparticle entanglement is a central problem in the
field of quantum information theory. Recently, this problem has received
significant attention for two main reasons: first, thanks to the hard work of
many experimentalists, multiparticle entanglement has been observed in ion traps
\cite{iontraps}, photon polarization \cite{photons}, and nitrogen-vacancy
centers in diamond \cite{nv}. Second, multiparticle entanglement has turned out
to be much more complex than two-particle entanglement. From a theorist's
perspective, this offers the possibility to work on mathematical challenges with
additional difficulties and joy.

One of these challenges is the question whether or not a given quantum state
contains genuine multiparticle entanglement. Despite many recent advances
\cite{hororeview,gtreview,seevinck,huber,bastian} and partial results, there is
no known general criterion. Progress on this is vital for experimentalists to
properly interpret their measurement results.

\begin{figure}[t]
\centering
\includegraphics[scale=0.5]{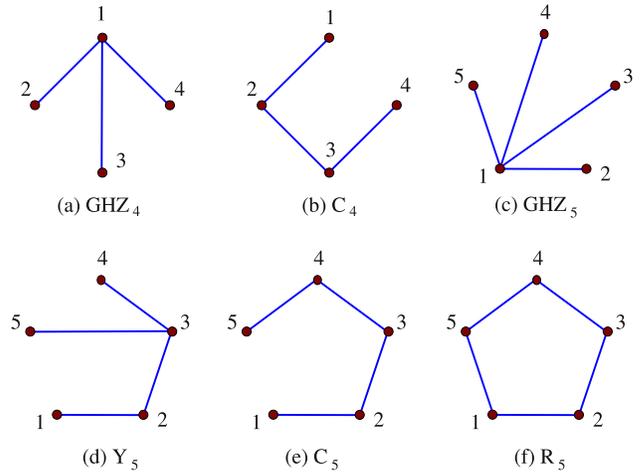}
\caption{The graphs of the states discussed in this paper. (a) and (c): 
star graphs corresponding to GHZ states. For GHZ-diagonal states with an
arbitrary number of qubits, the problem of detecting multiparticle entanglement
was already solved in Ref.~\cite{seevinck}. (b) The graph of the four-qubit
cluster state. Theorem 3 gives a necessary and sufficient criterion for
multipartite entanglement in states which are diagonal in the corresponding
graph-state basis. (d,e,f): five-qubit graph states
$\ket{Y_5},\ket{C_5},\ket{R_5}$. In section IV we determine the border of
separability when the graph diagonal generalizations of these states are mixed
with white noise. For graph-diagonal states corresponding to the $Y_5$-graph, we
also obtain a complete characterization of multiparticle entanglement (Theorem
10), which can be generalized to an arbitrary number of qubits (Theorem 12).}
\label{fig:thresh1}
\end{figure}

In this paper, we solve the problem of characterizing genuine multiparticle
entanglement for certain families of graph-diagonal states,
cf.~Fig.~\ref{fig:thresh1}. Graph states are multi-qubit states which are
extremely important for many aspects of quantum computation including quantum
error correction \cite{hein}. Graph-diagonal states are states which are
diagonal in 
the associated basis of this graph. Interest in this type of states is
physically motivated: they occur naturally upon the decoherence of pure graph
states \cite{graphdiagonal}, and, more importantly, any state can be brought
into graph-diagonal form by local operations \cite{graphdiagonal,
graphdiagonal2}. As local operations do not affect entanglement properties, this
means that if the corresponding graph-diagonal state is entangled the original
state was entangled. Thus, entanglement criteria for graph-diagonal states
produce entanglement criteria for general states. We note that a widely
discussed subset of graph-diagonal states are
Greenberger-Horne-Zeilinger-diagonal (GHZ diagonal) states. For these states the
characterization of genuine multiparticle entanglement has already been solved
\cite{seevinck}.

This paper is organized as follows: in Section~II  we introduce the relevant
notions of multiparticle entanglement and provide the definition and main
properties of graph states. In Section~III we specifically consider the
four-qubit cluster state and states diagonal in the corresponding basis. We
provide a necessary and sufficient criterion for genuine multiparticle
entanglement, and, for states without multiparticle entanglement, we provide an
explicit decomposition into biseparable states. In Section~IV we discuss
separability conditions for all five-qubit graph states mixed with white noise.
Again, we provide necessary and sufficient criteria for these families and
explicit decompositions when the states are separable.

In Section~V we relate our results to a recent approach to 
{characterize} multiparticle entanglement via so-called positive
partial-transpose (PPT) mixtures \cite{bastian}. Our result for the four-qubit
cluster state implies that this criterion is necessary and sufficient for the
four-qubit case. The question arises whether this is true in general. We argue
that this may not be the case. Nevertheless, in Section VI we discuss in detail
the five-qubit Y-shaped graph state for which we can prove that the method of
PPT mixtures does deliver a complete solution. In Section~VII we discuss
generalizations of the Y-shaped graph to an arbitrary number of qubits. Our
conclusions and a discussion of possible extensions of our work is presented in
Section~VIII. 

\section{Multiparticle entanglement and graph states}

In this section, we will review the basic notions of genuine multipartite
entanglement and graph states. Detailed presentations may be found in
Refs.~\cite{hororeview,gtreview,hein} and a reader familiar with these topics
may skip to the next section.

Let us start with the relevant definitions of multiparticle entanglement for
three particles. The generalization to more particles is straightforward. A pure
state is called fully separable, if it can be written in the form
$\ket{\psi^{\rm fs}}=\ket{a}\otimes \ket{b}\otimes\ket{c}$. A mixed state is
fully separable if it can be written as a convex combination of such fully
separable pure states, 
$
\vr^{\rm fs} = \sum_{k} p_k  \ketbra{\psi^{\rm fs}_k},
$
where the {coefficients $p_k$ form a probability distribution, i.e., $p_k \geq
0$ and $\sum_k p_k = 1.$}

A pure state is called biseparable if it is separable for some bipartition. For
example, the state $\ket{\psi^{\rm bs}}=\ket{a}\otimes\ket{\phi^{bc}}$ where
$\ket{\phi^{bc}}$ is a possibly entangled state of particles $B$ and $C$ is
biseparable for the $A|BC$-partition. Other bipartitions for three particles are
the $B|AC$- or $C|AB$-partition. A mixed state is biseparable if it can be
written as a sum of biseparable states:
\be
\vr^{\rm bs} = \sum_{k} p_k  \ketbra{\psi^{\rm bs}_k}.
\ee
Note that the states $\ket{\psi^{\rm bs}_k}$ may be biseparable for different
partitions. A state is genuine multipartite entangled if it is not biseparable.
Genuine multiparticle entanglement is typically the type of entanglement one
aims for in experiments \cite{gtreview}. Consequently, many aspect of
multparticle entanglement are under intensive research \cite{seevinck, huber,
bastian}. It is the  aim of this paper to derive necessary and sufficient
conditions of genuine multipartite entanglement for certain families of mixed
states known as graph-diagonal states.

Graph states are multi-qubit states defined as follows~\cite{hein}. Let $G$ be a
graph: a set of $N$ vertices corresponding to qubits with edges connecting them.
Some examples of graphs are shown in Fig.~\ref{fig:thresh1}. For each vertex
$i$, its neighbourhood $\NN(i)$ is the set of vertices connected to $i$ by an
edge. To define a graph state we associate a stabilizing operator $g_i$ to each
vertex $i$:
\begin{equation}
g_i:= X^{(i)}\bigotimes\nolimits_{j\in \NN(i)} Z^{(j)},
\label{stabdef}
\end{equation}
where $X^{(i)}, Y^{(i)}, Z^{(i)}$ denote the Pauli matrices
$\sigma_x,\sigma_y,\sigma_z,$ acting on the $i$-th qubit and the identity
operator on the rest of the qubits. The qubit index ${(i)}$ may be omitted
whenever there is no risk of confusion. The graph state $\ket{G}$ associated
with the graph $G$ is the unique $N$-qubit state fulfilling
\begin{equation}
g_i \ket{G}= \ket{G}, \mbox{ for } i=1,...,N. 
\label{graphdef}
\end{equation}
Hence, it is the unique $+1$ eigenstate to all stabilizing operators. Well known
examples for graph states are the GHZ states which correspond to the star shaped
graphs in Figs.~\ref{fig:thresh1}(a) and \ref{fig:thresh1}(c).

Alternatively, the graph describes a possible construction method of the graph
state. Starting with each qubit in the state
$\ket{x^+}=(\ket{0}+\ket{1})/\sqrt{2}$ one applies controlled phase-gates to all
connected qubits. This results is the graph state $\ket{G}.$ Note that the order
in which the phase gates are applied is irrelevant since the controlled
phase-gates commute. The construction method for graph states implies, for
instance, that the five-qubit linear cluster state [Fig.~\ref{fig:thresh1}(e)]
can be viewed as originating from a four-qubit cluster state
[Fig.~\ref{fig:thresh1}(b)] with an added fifth qubit. We will use this property
to prove separability of certain mixed states as follows: if one has a state
associated to the four-qubit cluster state which is known to be separable with
respect to some partition (say, $AB|CD$ or $12|34$, for definiteness), then the
state where the fifth qubit is added [as in Fig.~\ref{fig:thresh1}(e)] is still
separable with respect to the $AB|CDE$ partition.

The definition of a graph state can be extended to different eigenvalues of
$g_i$ and one may consider all the $2^N$ possible states $\ket{\Gamma_k}$ where
$g_i \ket{\Gamma_k}= a^{(i)}_k \ket{\Gamma_k}$ with eigenvalues $a^{(i)}_k=\pm
1.$ These vectors are all orthogonal and form the so-called graph state basis
for the Hilbert space of $N$ qubits.
The $\ket{\Gamma_k}$ are uniquely characterized by the $a^{(i)}_k$, so one
may also write $\ket{\Gamma_k}=\ket{a^{(1)}_k, a^{(2)}_k, ...}$. One can express
any of the states as $\ket{\Gamma_k}$:
\be
\ketbra{\Gamma_k} = \frac{1}{2^N}\prod_{i=1}^{N} (\eins + a^{(i)}_k g_k),
\label{expansion}
\ee
which will prove useful for our calculations.

In this paper, we focus on graph-diagonal states of the form
\be
\vr = \sum_k \lambda_k \ketbra{\Gamma_k},
\label{eq5}
\ee
where the parameters $\lambda_k$ form a probability distribution. This family of
states not only has nice mathematical properties but is important for physical
reasons as well. An arbitrary $N$-qubit state $\tilde \vr$  can always be
brought into a graph-diagonal form by local operations. Moreover, this so-called
depolarization map does not change the fidelities $\lambda_k =
\bra{\Gamma_k}\tilde \vr\ket{\Gamma_k}$ of the graph-basis states. Hence, if the
associated graph-diagonal state is entangled then the original $\tilde \vr$ was
entangled as well. For experiments, one can measure all the fidelities of the
graph states and consider the corresponding graph-diagonal state. Consequently,
studying graph-diagonal states has direct consequences for general states, and
many properties of graph-diagonal states have been studied in the past few years
\cite{graphdiagonal,graphdiagonal2}.

For our later discussion, we note that different graphs may lead to graph states
which only differ by a local unitary transformation, implying that their
entanglement properties are equivalent. The main graph transformation which
leaves the entanglement properties invariant is the so-called local
complementation. Local complementation acts on a graph as follows: for a vertex
$i$ invert its neighbourhood $\NN(i)$. This means that vertices in the
neighbourhood that were connected become disconnected and vice versa. For
example, local complementation on qubit 1 of graph in Fig.~1(a) will transform
it into a graph that is fully connected. Similarly, local complementation of
qubit 2 in Fig.~1(b) will connect qubits 1 and 3. Modulo local complementation,
there are only two independent four-qubit graphs and four five-qubit graphs all
of which are displayed in Fig.~\ref{fig:thresh1}.

As an example of a graph state, let us consider the graph in
Fig.~\ref{fig:thresh1}(b). The stabilizing operators of this graph are
\be
g_1 = X Z 1 1, 
\;\;
g_2 = Z X Z 1,
\;\;
g_3=1 Z X Z,
\;\;
g_4= 1 1 Z X,
\label{stabs}
\ee
and the associated graph state is the so-called cluster state $\ket{C_4}.$ After
a local transformation, 
the cluster state can also be written in the common representation 
\cite{hein}
\be
\ket{C_4'}=\frac{1}{2}(\ket{0000}+\ket{0011}+\ket{1100}-\ket{1111}),
\ee
given in the computational basis. We will work, however, in the basis defined by
the stabilizers in 
Eq.~(\ref{stabs}). The cluster state $\ket{C_4}$ is an eigenstate of the $g_i$
with $+1$ eigenvalue. 
As indicated above, we can write 
$\ket{C_4}$ as $\ket{\!+\!+\!+\!+}$, where the qubits are written in the order
$ABCD$.
Similarly, the other states in the basis can be denoted 
by $\ket{\!+\!+\!+\!-}, ..., \ket{\!-\!-\!-\!-}$. We write these states in
general as $\ket{ijkl}$ 
with $i,j,k,l \in \{+, -\}.$ When using this notation, we denote with 
$\bar{i}, \bar{j}, \bar{k},\bar{l}$  the opposite signs to $i,j,k,l$; such that
$i \bar{i} = -1$ 
etc. We also find it convenient to abbreviate projectors as
$\ketbrapro{ijkl}\equiv\ketbra{ijkl}.$ Finally, 
let us recall that the signs of all the states in the graph state basis can be
changed 
by applying the Pauli matrix $Z$ to the corresponding qubits. As these are local
operations they 
do not change the states' entanglement properties.

\section{Cluster-diagonal states of four qubits}

In this section, we will derive a necessary and sufficient criterion for the
presence of genuine multipartite entanglement in cluster-diagonal states of four
qubits. Before proving our main result, we need two lemmata. The first one
characterizes a set of entanglement witnesses for genuine multipartite
entanglement, while the second one identifies a large class of biseparable
quantum states that will simplify the search for biseparable decompositions.

\noindent
{\bf Lemma 1.}
The observables
\bea
\WW_1 &= &\frac{\eins}{2} - \ketbra{C_4} - \frac{1}{2} \frac{\eins -
g_1}{2}\frac{\eins - g_4}{2}
\nonumber
\\
&=& \frac{\eins}{2} - \ketbrapro{\!+\!+\!+\!+} - \frac{1}{2}
\sum_{ij}\ketbrapro{\!- ij -},
\nonumber
\\
\WW_2 &= & \frac{\eins}{2} - \ketbrapro{\!+\!+\!+\!+} - \ketbrapro{\!- \alpha
\beta -},
\label{c4witnesses}
\eea
are entanglement witnesses for genuine multipartite entanglement. That is, ${\rm
Tr}(\vr \WW_k) <0$
implies the presence of genuine multipartite entanglement in $\vr.$ This holds
for arbitrary signs $\alpha,\beta$
in $\WW_2.$

{\it Proof.}
It was proven in Ref.~\cite{bastian} that $\WW_1$ is a witness. The fact that
$\WW_2$ is a witness can be demonstrated in a similar way. It suffices to show
that $(\WW_2)^{T_M} \geq 0$ for all possible bipartitions $M.$ The operators
$(\WW_2)^{T_M}$ are diagonal in the graph state basis. Thus, it is enough to
show that $\bra{ijkl}(\WW_2)^{T_M}\ket{ijkl}\geq 0$ holds for all elements of
the graph state basis. This, however, is a direct consequence of Lemma 2 and 
Lemma 3 in the Appendix of Ref.~\cite{bastian}.
\qed

\noindent
{\bf Lemma 2.} The quantum states
\be
\sigma = \frac{1}{2}(\ketbrapro{ijkl} + \ketbrapro{\alpha \beta \gamma \delta})
\ee
are biseparable, unless $i\neq \alpha$ and $l \neq \delta$ both hold at the same
time.

{\it Proof.} First, we note that if $i\neq \alpha$ and $l \neq \delta$ the state
is definitely not biseparable, since it is detected by a witness of the type
$\WW_2$ (and also by $\WW_1$). Now, we show explicitly that all other states are
biseparable. We can assume without loss of generality that
$\ket{ijkl}=\ket{\!+\!+\!+\!+}$  since any $\ket{ijkl}$ can be transformed into
$\ket{\!+\!+\!+\!+}$ by local transformations and we neglect the normalization
of $\sigma$. The first example is presented in great detail so as to demonstrate
our methodology.

(a) Consider the state $\sigma = \ketbrapro{\!+\!+\!++} +
\ketbrapro{\!-\!+\!++}.$ There are two ways to see that $\sigma$ is biseparable
with respect to the $A|BCD$ partition and we will discuss both of them, in order
to illustrate the different methods. 

(a1) The first method starts with the fact that for two qubits any mixture of
two Bell states with equal weight (e.g., $\eta = \ketbra{\Phi^+} +
\ketbra{\Phi^-}$) is separable \cite{horoalt}. The graph corresponding to a Bell
state is the connected two-qubit graph. The four-qubit state $\sigma$ can be
considered as a separable mixture of the two Bell states $\ketbrapro{\!++} +
\ketbrapro{\!-+}$ on the first two qubits $AB$, where  the qubits $CD$ have been
subsequently added to $B$ via some local interaction. Clearly, the state
$\sigma$ remains biseparable between $A$ and the rest of the qubits.

(a2) The second method uses Eq.~(\ref{expansion}) to write
\begin{align}
\sigma &\sim (\eins +g_2)(\eins +g_3)(\eins+g_4)
\nonumber
\\
\label{eq:help1}
&\sim (\eins + ZXZ1) (\eins + 1ZXZ) (\eins + 11ZX)
\nonumber
\\
&\sim \underbrace{(\eins+Z111)}_{\sim \ketbra{0}}
\underbrace{(\eins+g_2^{\rm red})(\eins+g_3)(\eins+g_4)}_{\rm stabilizer \;\;
state \;\;on \;\; qubits \;\; 2,3,4 }
\nonumber
\\
&
+
\underbrace{(\eins-Z111)}_{\sim \ketbra{1}}
\underbrace{(\eins-g_2^{\rm red})(\eins+g_3)(\eins+g_4)}_{\rm stabilizer \;\;
state \;\;on \;\; qubits \;\; 2,3,4 },
\end{align}
where $g_2^{\rm red}= 1XZ1$ denotes the restriction of the stabilizer $g_2$ to
the qubits 2,3,4. In this form the state is clearly biseparable, since it is
written as a sum of two terms, which are both biseparable with respect to the
$A|BCD$-partition. This rewriting is possible, since in the expansion $\sigma 
\sim (\eins +g_2)(\eins +g_3)(\eins+g_4)$ only the identity and one of the Pauli
matrices (here: $Z$) occur on the first qubit, cf. Eq.~(\ref{eq:help1}). This
statement holds for any Pauli operator. With these two methods in hand we now
prove the other states, $\sigma$, are also biseparable. 

(b) We consider $\sigma = \ketbrapro{\!+\!+\!+\!+} + \ketbrapro{\!+\!-\!+\!+}.$
Using (a1) this is clearly separable with respect to the $A|BCD$ partition, but
it is also separable with respect to the $B|ACD$-partition, as can be seen using
the idea of (a2).

(c) The state $\sigma = \ketbrapro{\!+\!+\!+\!+} + \ketbrapro{\!-\!-\!+\!+}$ is
biseparable with respect to the $A|BCD$-partition according to (a1).

(d) The state $\sigma = \ketbrapro{\!+\!+\!+\!+} + \ketbrapro{\!-\!+\!-\!+}
\sim
(\eins + g_1g_3) (\eins + g_2)(\eins + g_4)
$
is biseparable with respect to the $B|ACD$ partition, as can be seen using (a2).

(e) The state $\sigma = \ketbrapro{\!+\!+\!+\!+} + \ketbrapro{\!+\!-\!-\!+}$ can
be shown to be biseparable using the method of (a1) with qubits $B$ and $C$ as
the Bell pair. Consequently, it is separable with respect to the
$AB|CD$-partition.

(f) Finally, we consider $\sigma = \ketbrapro{\!+\!+\!+\!+} +
\ketbrapro{\!-\!-\!-\!+} \sim (\eins + g_1 g_2 + g_2 g_3+ g_1 g_3)(\eins +
g_4).$ First, using the method of (a2) one can directly calculate that this
state is separable with respect to the $B|ACD$-partition. However, one can also
apply the method of (a1): On the first three qubits, one can consider the state
$\sigma = \ketbrapro{\!+\!++} + \ketbrapro{\!-\!- -}.$ This corresponds to a
mixture of two three-qubit GHZ states, and it is known that such mixtures are
always biseparable \cite{seevinck}. For $\sigma$ only one qubit is added similar
to (a1), so $\sigma$ has to be biseparable, too. Up to symmetries these are all
the relevant cases. 
\qed

We can now formulate and prove our main result. We denote the fidelities of the 
graph basis states as $F_{++++}=\bra{\!+\!+\!+\!+}\vr\ket{\!+\!+\!+\!+}$
etc.~We 
can then state:

{\bf Theorem 3.} A cluster-diagonal  four-qubit state $\vr$ is biseparable, if
and only if for all indices $\alpha, \beta,\gamma,\delta$
\begin{align}
2 & F_{\alpha \beta \gamma \delta} 
\leq \sum_{i,j} F_{\alpha ij \delta} 
+ \sum_{i,j} F_{\bar{\alpha} ij \delta}+
\sum_{i,j} F_{\alpha ij \bar{\delta}}
\label{condition1}
\end{align}
holds and for all indices $\alpha, \beta, \gamma, \delta, \mu, \nu$ the
inequalities
\begin{align}
2   F_{\alpha \beta \gamma \delta} + 2 F_{\bar{\alpha}\mu \nu \bar{\delta}} 
\leq &
\sum_{i,j} F_{\alpha ij \delta} 
+ \sum_{i,j} F_{\bar{\alpha} ij \delta}
\nonumber
\\
&+
\sum_{i,j} F_{\alpha ij \bar{\delta}}
+
\sum_{i,j} F_{\bar{\alpha} ij \bar{\delta}}
\label{condition2}
\end{align}
are satisfied.

Before proving this result, let us interpret the conditions in
Eqs.~(\ref{condition1}, \ref{condition2}). In light of Lemma~2,
Eq.~(\ref{condition1}) compares the weight of the state $\ketbrapro{\alpha \beta
\gamma \delta}$ with the sum of the weights of all other states, which can be
used to build a biseparable pair with $\ketbrapro{\alpha \beta \gamma \delta}$.
If the overall state is biseparable the first weight has to be smaller than the
other weights, otherwise a decomposition with the methods of Lemma~2 cannot be
found. The condition Eq.~(\ref{condition2}) then compares the weights of two
states, $\ketbrapro{\alpha \beta \gamma \delta}$ and  $\ketbrapro{\bar{\alpha}
\beta \gamma \bar{\delta}}$ (which, according  to Lemma 2, do not constitute a
separable pair) with all other weights. Using the normalization of the state,
Eq.~(\ref{condition2}) can be rephrased as $F_{\alpha \beta \gamma \delta} +
F_{\bar{\alpha}\mu \nu \bar{\delta}} \leq 1/2$, which has a natural meaning: if
the weight of one ``inseparable'' pair exceeds all other weights, then the state
cannot be separable.

{\it Proof.} 
We will use a shorthand notation for the sums such that Eq.~(\ref{condition1}),
$2 F_{++++} \leq \sum_{ij} F_{+ij+} + \sum_{ij} F_{-ij+} +\sum_{ij} F_{+ij-}$,
is abbreviated as $2 F_{++++} \leq \sum_{++}  + \sum_{-+}  +\sum_{+-}.$

We first have to show that if one of the conditions in Eqs.~(\ref{condition1},
\ref{condition2}) is violated, then $\vr$ is genuinely multipartite entangled.
This follows directly from Lemma 1 since the conditions (\ref{condition1},
\ref{condition2}) are nothing but a rewriting of ${\rm Tr}(\WW_k\vr)\geq 0.$

It remains to show that a state is biseparable, if Eqs.~(\ref{condition1},
\ref{condition2}) hold. Clearly, this is the difficult part. Our proof is split
into four cases:

{\it Case 1 ---} Let us first assume that the state $\vr$ acts only on the
four-dimensional space spanned by the vectors $\ket{\!-\!ij-}$ and that the
relevant four fidelities fulfill $F_{- \alpha \beta -} \leq F_{-\alpha
\bar{\beta}-}+ F_{-\bar{\alpha}\beta-}+ F_{-\bar{\alpha}\bar{\beta}-}$ for all
$\alpha, \beta$. Then, the state $\vr$ is separable with respect to the $AB|CD$
partition. The reason is the following: a mixture of two-qubit Bell states,
$\sigma = \lambda_{++} \ketbra{\Phi^+} + \lambda_{+-} \ketbra{\Phi^-} +
\lambda_{-+} \ketbra{\Psi^+} + \lambda_{--} \ketbra{\Psi^-}$, is easily seen to
be separable iff $\lambda_{\alpha \beta} \leq \lambda_{\alpha \bar{\beta}}+
\lambda_{\bar{\alpha}\beta} + \lambda_{\bar{\alpha}\bar{\beta}}$ for all
$\alpha, \beta$ \cite{horoalt}. The four-qubit state $\vr$ is nothing but a
mixture of such Bell states between $B$ and $C$, with qubits $A$ and $D$ added
[see also case (a1) in the proof of Lemma 2].

{\it Case 2 ---} Now we assume that equality holds for one of the conditions of
Eq.~(\ref{condition1}). Without loss of generality, we assume that $2 F_{++++} =
\sum_{++}  + \sum_{+-}  +\sum_{-+}$ while the other conditions in
Eqs.~(\ref{condition1}, \ref{condition2}) are fulfilled, but not necessarily
with equality.

In this case, Eq.~(\ref{condition2}) becomes $F_{- \alpha \beta -} \leq
F_{-\alpha \bar{\beta}-}+ F_{-\bar{\alpha}\beta-}+
F_{-\bar{\alpha}\bar{\beta}-}$, the same relation discussed in Case 1. If we
consider now the projection $\vr^{\rm R}$ of the original state $\vr$ on the
four-dimensional space spanned by the vectors $\ket{\!-\!ij-},$ then it is clear
that this state  $\vr^{\rm R}$ is separable according to Case 1. It remains to
show that the orthogonal part $\vr -\vr^{\rm R} $ is separable too. For this
part we have  $F_{++++} = F_{+--+}+ F_{++-+}+F_{+-++} + \sum_{+-} +\sum_{-+,}$
so it can be directly decomposed with the help of Lemma~2,
by using all possible combinations of the type \linebreak  
$\ketbrapro{\!+\!+\!++} + \ketbrapro{\alpha \beta \gamma \delta}$.
This finishes the proof of Case 2.

{\it Case 3 ---} In this case we assume that equality holds for one of the
conditions of Eq.~(\ref{condition2}): $2 F_{++++} + 2 F_{----} = \sum_{++}  +
\sum_{+-}  +\sum_{-+}+ \sum_{--}.$ The other inequalities in
Eqs.~(\ref{condition1}, \ref{condition2}) are satisfied, but not necessarily
with equality.
Rewriting, gives us $(2 F_{++++} - \sum_{++}) + (2 F_{----}-\sum_{--}) = 
\sum_{+-}  +\sum_{-+}.$ Using this together with the inequalities given by
Eq.~(\ref{condition1}) we can also deduce the conditions $2 F_{----} \geq
\sum_{--}$ and $2 F_{++++} \geq \sum_{++}$. 

Now, we can decompose $\vr$ as follows: Consider the space spanned by
$\ket{\!-\!ij-},$  and a state $\sigma^-$ with $F'_{----}:=\sum_{--}-F_{----} 
= F_{--+-}+F_{-+--}+F_{-++-}$ and $F'_{-ij-}:=F_{-ij-}$ otherwise. 
This state is separable according to Case 1, since 
$F'_{----} =  F'_{--+-}+F'_{-+--}+F'_{-++-}$. 
The restriction $\vr^{\rm R}$ of $\vr$ onto the four-dimensional subspace is now
given by $\vr^{\rm R}= \sigma^- + (F_{----} -F'_{----} ) \linebreak
\ketbrapro{\!-\!-\!--} = \sigma^- + (2 F_{----}-\sum_{--}
)\ketbrapro{\!-\!-\!--}.$ We can make a similar construction on the space
spanned by $\ket{\!+\!ij+}$ with a separable state $\sigma^+$. A projector onto
$\ket{\!+\!+\!+\!+}$ with weight $(2 F_{++++} - \sum_{++})$ will remain.

Therefore, we can decompose $\vr$ into the two separable states $\sigma^-$
and $\sigma^+$ on the four-dimensional spaces and a remaining state $\eta.$
The state $\eta$ has only two contributions on the two four-dimensional spaces,
which have the fidelities $F_{++++}^{\eta}=2 F_{++++} - \sum_{++}$
and $F_{----}^{\eta}=2 F_{----}-\sum_{--}.$ From our assumption, it follows
that $\eta$ fulfils $ F_{++++}^{\eta} + F_{----}^{\eta} =  \sum_{+-} 
+\sum_{-+}$. This remaining state  $\eta$ can then be decomposed using states of
the form $\sigma=\ketbrapro{\!-\!-\!-\!-}+\ketbrapro{\!-\!kl+}$ and
$\sigma=\ketbrapro{\!-\!-\!-\!-}+\ketbrapro{\!+\!kl-}$ etc.

{\it Case 4 ---} Let us finally discuss the case where equality holds for none
of the conditions of Eqs.~(\ref{condition1}, \ref{condition2}). We consider the
state
\be
\vr^{\rm new}= \vr - \varepsilon \sigma,
\ee
where $\sigma$ is one of the separable states from Lemma 2. Since $\vr=\vr^{\rm
new}+\varepsilon \sigma$ the state $\vr$ is separable, if $\vr^{\rm new}$ is
separable and positive.

The idea is to choose possible biseparable states $\sigma$ and subtract them
step by step such that $\vr^{\rm new}$ remains positive. Note that during these
subtractions, the inequalities (\ref{condition1}, \ref{condition2}) become
tighter. But one does not have to worry that they become violated: If they
become violated, at some point equality must hold in one of the two
Eqs.~(\ref{condition1},\ref{condition2}) first, while the other conditions still
hold. This means that at this point $\vr^{\rm new}$ (and hence $\vr$) is
separable, according to Cases 2 and 3.

What can be achieved with the iterative subtractions? First, by subtracting the
biseparable states \mbox{$\sigma= \ketbrapro{\!+\!ij+}+\ketbrapro{\!+\!kl+}$}
one can set three of the $F_{+\alpha\beta+}$ to zero. Similarly, in each of the
sets $\{F_{+\alpha\beta-}\}$, $\{F_{-\alpha\beta+}\}$ and
$\{F_{-\alpha\beta-}\}$ three fidelities can be made to vanish, such that
overall only four $F_{ijkl}$ are nonzero. The structure of the fidelities is now
such that all the sums in Eqs.~(\ref{condition1}, \ref{condition2})
contain only a single term. Then, however, Eq.~(\ref{condition2}) must either be
violated for some set of indices, or equality must hold.
\qed


Using this theorem we can determine that cluster states mixed with white noise,
$\vr(p) =p \ketbra{C_4} +(1-p){\eins}/{16}$, are entangled iff $p>5/13.$ This
confirms a numerically established threshold from Ref.~\cite{bastian}.

Furthermore, the theorem demonstrates that for cluster-diagonal states there are
effectively only two entanglement witnesses, namely the ones from Lemma 1. It is
interesting to compare this with the results of Ref.~\cite{seevinck}, where a
necessary and sufficient criterion for GHZ diagonal states was found. This
criterion can be interpreted in the sense that for GHZ diagonal states (of an
arbitrary number of qubits) only one entanglement witness is relevant, namely
$\WW=\eins/2 - \ketbra{GHZ_N}.$ For cluster states, the witness $\WW=\eins/2 -
\ketbra{C_4}$ is not optimal, since both of the witnesses in Lemma 1 are better.
One can expect that for more complicated graph states of more qubits, a
significant higher number of witnesses is  relevant, hence a complete
classification becomes difficult.

\section{Five-qubit graph states}

In this section, we derive optimal criteria for all five-qubit graph states
mixed with white noise. Doing this demonstrates that the witnesses obtained with
the PPT approach of Refs.~\cite{bastian, bastian2} are optimal. In the next
section, however, we will argue that the success of the PPT approach in finding
optimal witnesses might be specific to these states. Nevertheless, we do present
full solution of the cluster state $Y_5$ in Section VI, cf. Theorem~10.

\subsection{The state $Y_5$}

For the five-qubit $Y_5$ state [see Fig.~1(d)] mixed with white 
noise we demonstrate:

{\bf Proposition 4.} The state
\be
\vr(p) =p \ketbra{Y_5} +(1-p)\frac{\eins}{32}
\ee
is genuine multipartite entangled if and only if $p>9/25=0.36.$

{\it Proof.} 
First, for the case that $p>9/25$ the state $\vr(p)$ is detected 
by the witness \cite{bastian2}
\begin{align}
&\WW_{Y5} = \frac{\eins}{2} - \ketbra{Y_5} 
-
\frac{1}{16}
\big[(\eins-g_1)(\eins-g_4)(\eins+g_5) 
\nonumber
\\
&+
(\eins-g_1)(\eins+g_4)(\eins-g_5) 
+
(\eins-g_1)(\eins-g_4)(\eins-g_5) 
\big]
\label{y5wit}
\end{align}
and, hence, genuine multipartite entangled.

In the other direction, we first have to identify the separable states as we did
in Lemma 2. In fact, for many states this lemma can be directly generalized. For
instance, the state $\sigma=\ketbrapro{\!+\!+\!+\!++}+\ketbrapro{ijk\!++}$ is
biseparable, since for the four-qubit cluster state
$\sigma'=\ketbrapro{\!+\!+\!+\!+}+\ketbrapro{ijk+}$ is separable, and the fifth
qubit is added as in case (a1) in the proof of Lemma 2. In fact, the only
combinations which are {\it not} separable are of the form
$\chi_1=\ketbrapro{\!+\!+\!+\!++}+\ketbrapro{\!-\!jk\!-+}$ and
$\chi_2=\ketbrapro{\!+\!+\!+\!+ +}+\ketbrapro{\!-\!jkl\!-}$. Note that the state
$\sigma=\ketbrapro{\!+\!+\!+\!++}+\ketbrapro{+jk\!-\!-}$ is biseparable, because
it can be considered as a separable four-qubit GHZ state {on BCDE} where one
qubit is added [see case (f) in the proof of Lemma 2].

The state at the critical value  of $p$ is 
\be
\vr\sim 19 \ketbrapro{\!+\!+\!+\!+ +} + \sum_{ijklm\neq+++++}\ketbrapro{ijklm},
\ee 
and it remains to show that this state is separable. First, the state 
\begin{align}
&\vr'= 19 \ketbrapro{\!+\!+\!+\!+ +} +
\big[\sum_{ijklm\neq+++++}\ketbrapro{ijklm}  
\nonumber
\\
&-\sum_{ij}\ketbrapro{\!-\!ij\!-\!+}- \sum_{ij}\ketbrapro{\!-\!ij\!+\!-} -
\sum_{ij}\ketbrapro{\!-\!ij\!-\!-} \big]
\end{align}
is biseparable, since in the sums in the brackets exactly 19 terms remain, and a
decomposition with $\sigma$ 
from above is then straightforward. The remaining term
$\vr - \vr' = \sum_{ij}\ketbrapro{\!-\!ij\!-\!+}+
\sum_{ij}\ketbrapro{\!-\!ij\!+\!-} + \sum_{ij}\ketbrapro{\!-\!ij\!-\!-}$
is also clearly separable, since the sum of any two of the occurring states is
separable.
\qed

\subsection{The linear cluster state $C_5$}

For the five-qubit linear cluster $C_5$ state [see Fig. 1(e)] mixed with white
noise the threshold is the same as for the $Y_5$ state:

{\bf Proposition 5.} The state
\be
\vr(p) =p \ketbra{C_5} +(1-p)\frac{\eins}{32}
\ee
is genuine multipartite entangled if and only if $p>9/25=0.36.$

{\it Proof.} First, the witness
\begin{align}
&\WW_{C5}= \frac{\eins}{2} - \ketbra{C_5} 
-
\frac{1}{32}
\big[4  (\eins-g_1)(\eins-g_5) \nonumber
\\
&+
(\eins+g_1)(\eins-g_2)(\eins-g_5) 
+
(\eins-g_1)(\eins-g_4)(\eins+g_5) 
\big]
\label{c5wit}
\end{align}
detects the state for $p>9/25,$ proving one part of the claim \cite{bastian2}.

For the other direction, we have to again identify the biseparable states.
First, in a generalization of Lemma~2, states of the form
$\sigma=\ketbrapro{\!+\!+\!+\!++}+\ketbrapro{ijklm}$ are separable, unless they
are of the form $\chi_1=\ketbrapro{\!+\!+\!+\!+ +} +\ketbrapro{\!-\!jk\!-+},$
$\chi_2=\ketbrapro{\!+\!+\!+\!+ +} +\ketbrapro{\!+\!-\!jk-},$ or
$\chi_3=\ketbrapro{\!+\!+\!+\!++}+\ketbrapro{\!-\!jkl-}.$ There are 16 terms of
this type which are not biseparable.

The state at $p=9/25$ is, up to normalization, given by $\vr = 19
\ketbrapro{\!+\!+\!+\!++} + \sum_{ijklm\neq+++++}\ketbrapro{ijklm}.$
Generalizing Lemma 2 we can subtract many pairs of terms such that what remain
is to show that 
\begin{align}
\vr'= & 4  \ketbrapro{\!+\!+\!+\!++} + \sum_{ij}\ketbrapro{\!-ij-\!+}
\nonumber
\\
&
+ \sum_{ij}\ketbrapro{\!+\!-ij-} + \sum_{ijk}\ketbrapro{\!-\!ijk-}
\label{remainingstate}
\end{align}
is separable. To do this let us consider the four states
\bea
\eta_1 &=& \ketbrapro{\!+\!+\!+\!++} + \ketbrapro{\!+\!-\!+\!+-} 
+\ketbrapro{\!-\!+\!+\!-+} 
\nonumber \\ &&
+\ketbrapro{\!-\!-\!+\!--},
\nonumber \\
\eta_2 &=& \ketbrapro{\!+\!+\!+\!++} + \ketbrapro{\!+\!-\!+\!--}
+\ketbrapro{\!-\!-\!+\!-+} 
\nonumber \\ && +\ketbrapro{\!-\!+\!+\!+-},
\nonumber \\
\eta_3 &=& \ketbrapro{\!+\!+\!+\!++} + \ketbrapro{\!+\!-\!-\!+-}
+\ketbrapro{\!-\!-\!-\!-+} 
\nonumber \\ && +\ketbrapro{\!-\!+\!+\!--},
\nonumber \\
\eta_4 &=& \ketbrapro{\!+\!+\!+\!++} + \ketbrapro{\!+\!-\!-\!--}
+\ketbrapro{\!-\!+\!-\!-+} 
\nonumber \\ && +\ketbrapro{\!-\!-\!+\!+-}.
\label{lc5discussion}
\eea
The state $\eta_1$ is separable for the following reason: it is 
known that the four-qubit Smolin state 
$\sigma = 
\ketbra{\Phi^+}_{AB} \otimes \ketbra{\Phi^+}_{A'B'}+
\ketbra{\Phi^-}_{AB} \otimes \ketbra{\Phi^-}_{A'B'}+
\ketbra{\Psi^+}_{AB} \otimes \ketbra{\Psi^+}_{A'B'}+
\ketbra{\Psi^-}_{AB} \otimes \ketbra{\Psi^-}_{A'B'}$ 
is separable with respect to  the $AA'|BB'$ partition \cite{smolin}. 
The state $\eta_1$  is simply a Smolin state between the qubits $ABDE$, where
the qubit $C$ has been added [see case (a1) in Lemma 2]. Therefore, it is
separable with respect to the $AE|BCD$-partition. Similarly, $\eta_2$ is a
Smolin state up to local unitary operations and therefore separable with respect
to the same partition.

It can be directly verified that the state $\eta_3$ is PPT with respect to the
$BD|ACE$ partition. This implies separability via the following argument: for
the considered partition, $\eta_3$ is acting on a $4 \times 8$ (effectively $4
\times 4$) space. The PPT entangled states in this scenario have at least a rank
of five \cite{horobounddim}. Hence, $\eta_3$, which is of rank four, must be
separable with respect to the partition \footnote{Alternatively, one can see the
separability of $\eta_3$ as follows: applying local complementation on qubit 2
and then on qubit 1 exchanges qubits 1 and 2. Similarly, a local complementation
first on qubit 4 and then on qubit 5 exchanges qubits 4 and 5. The signs of the
states $\ket{ijklm}$ in the graph-state basis are not invariant under these
transformations. Applying the rules of a local complementation \cite{hein}, a
complementation on qubit $a$ flips the signs in the neighbourhood $\NN(a)$ if
and only if the sign on $a$ is $-1$. With this rule, one sees that after a
complementation on the qubits 2, then 1, then 4, then 5 the state is like a
Smolin state between the qubits ABDE, and the  qubit C is connected to the
qubits A and E, so it is separable with respect to the $BD|ACE$ partition. The
same argument can be applied to $\eta_4$.}. Similarly, $\eta_4$ is separable
with respect to the $BD|ACE$-partition.

So we can write
\be
\vr'=\sum_{k=1}^4 \eta_k + \sum_{ij}\ketbrapro{\!-\!i\!-\!j-}
\ee
where the sum of the remaining four projectors is clearly separable according to
Lemma 2. This finishes the proof.
\qed

\subsection{The ring cluster state $R_5$}

For the five-qubit ring cluster state mixed with white noise the separability
problem can be solved as follows:

{\bf Proposition 6.} The state
\be
\vr(p) =p \ketbra{R_5} +(1-p)\frac{\eins}{32}
\ee
is genuine multipartite entangled if and only if $p>7/19 \approx 0.368.$

{\it Proof.}  Due to the symmetry of this state, it is convenient for our
discussion to define $\TT(x)$ as the sum over all five translations of the term
$x$, corresponding to a rotation of the ring graph. The witness for the state is
then given in \cite{bastian2}:
\bea
\WW_{R5} &=& 3\times \big[\TT(\ketbrapro{\!+\!+\!+\!+-}) +
\TT(\ketbrapro{\!+\!+\!-\!+-})
\nonumber
\\
&& + \TT(\ketbrapro{\!+\!+\!-\!--})\big]
\nonumber
\\
& +&
\big[\TT(\ketbrapro{\!+\!+\!+\!--})+ \TT(\ketbrapro{\!+\!-\!+\!--}) 
\nonumber
\\
&& 
+ \TT(\ketbrapro{\!+\!-\!-\!--})\big]
\nonumber
\\
&-&
\ketbrapro{\!-\!-\!-\!--} - 3 \times \ketbrapro{\!+\!+\!+\!++}.
\label{r5wit}
\eea
This witness detects the entanglement in the state for $p>7/19$, proving one
direction of the claim.

For the other direction, we have to identify separable states. First, states
like 
$\sigma_1=\ketbrapro{\!+\!+\!+\!++}+\ketbrapro{\!+\!+\!+\!+-}$ 
$\sigma_2=\ketbrapro{\!+\!+\!+\!++}+\ketbrapro{\!+\!+\!-\!+-}$ 
and
$\sigma_3=\ketbrapro{\!+\!+\!+\!++}+\ketbrapro{\!+\!+\!-\!--}$
are clearly separable in analogy to Lemma~2: $\sigma_1$ is separable in
analogy to case (a2), $\sigma_2$ and $\sigma_3$ can be considered as 
states on the qubits $BCDE$ which are separable with respect to the 
$D|BCE$ partition [cases (d) and (f) in Lemma 2], where the qubit $A$ 
is added by a local transformation.
 Furthermore, the states
\bea
\eta_1 &=& \ketbrapro{\!+\!+\!+\!++} + \ketbrapro{\!+\!-\!-\!++}
+\ketbrapro{\!-\!-\!+\!-+} 
\nonumber \\ 
&&+\ketbrapro{\!-\!+\!-\!-+},
\nonumber \\
\eta_2 &=& \ketbrapro{\!+\!+\!+\!++} + \ketbrapro{\!-\!-\!+\!++}
+\ketbrapro{\!+\!+\!-\!-+}
\nonumber \\ 
&& +\ketbrapro{\!-\!-\!-\!-+},
\nonumber \\
\eta_3 &=& \ketbrapro{\!+\!+\!+\!++} + \ketbrapro{\!-\!-\!+\!++}
+\ketbrapro{\!-\!+\!-\!--}
\nonumber \\ 
&& +\ketbrapro{\!+\!-\!-\!--},
\nonumber \\
\eta_4 &=& \ketbrapro{\!+\!+\!+\!++} + \ketbrapro{\!-\!-\!+\!-+}
+\ketbrapro{\!+\!-\!-\!+-} 
\nonumber \\ &&+\ketbrapro{\!-\!+\!-\!--}.
\eea
are also separable. The state $\eta_1$ is separable with respect to the
$BC|ADE$-partition, as can be seen from the separability properties of the
Smolin state (similar to the state $\eta_1$ defined for the linear cluster state
$C_5$ above). $\eta_2$ is PPT with respect to the $BC|ADE$-partition, and hence
separable (due to a rank argument as in the proof of Proposition 5.). The
separability of $\eta_3$ (and $\eta_4$) can be inferred from their being PPT
with respect to the $CE|ABD$ (and $AC|BDE$) partition.

The state at $p=7/19$ is given by $\vr = 59  \ketbrapro{\!+\!+\!+\!++} + 3
\times \sum_{ijklm\neq+++++}\ketbrapro{ijklm}$ which can be written as
\be
\vr = 3 \times \sum_{k=1}^3 \TT(\sigma_k) + \frac{14}{20} \times \sum_{k=1}^4
\TT(\eta_k) + \vr'
\ee
with
\bea
\vr'&=& \frac{1}{5} \TT(\ketbrapro{\!+\!+\!+\!--}) + \frac{1}{5}
\TT(\ketbrapro{\!+\!-\!+\!--})
\nonumber
\\
&&+
\frac{1}{5} \TT(\ketbrapro{\!+\!-\!-\!--}) + 3 \ketbrapro{\!-\!-\!-\!--}.
\eea
This state, however, can directly be decomposed in terms of the $\sigma_k$ with
all signs inverted.
\qed

\section{Connection with the theory of PPT mixtures}
In order to place our results within a wider framework, we discuss possible
connections with the theory of PPT mixtures, introduced in Refs.~\cite{bastian,
bastian2}. In these papers, the following approach to characterize multiparticle
entanglement has been proposed: instead of considering biseparable states of the
form
\be
\vr^{\rm bs} = \sum_{k} p_k  \ketbra{\psi^{\rm bs}_k},
\ee
where the {states} $\ketbra{\psi^{\rm bs}_k}$ are separable with respect to some
partition, consider states of the form
\be
\vr^{\rm pmix} = \sum_{k} p_k  \vr^{\rm ppt}_k,
\ee
with states $\vr^{\rm ppt}_k$ that have a positive partial transpose (PPT) with
respect to some bipartition. Such states are called PPT mixtures. Since
separable states are also PPT, the set of biseparable states is a subset of the
set of PPT mixtures. Consequently, proving that a state is not a PPT mixture
implies genuine multiparticle entanglement.

The advantage of this approach is that the set of PPT mixtures can be
characterized much more easily than the set of biseparable states. For instance,
for a small number of (up to seven) qubits one can directly decide whether a
state is a PPT mixture via the method of semidefinite programming
\cite{pptmixture}. Moreover, witnesses detecting states that are not PPT
mixtures can be derived analytically. {This has been done for many types of
graph states} in Ref.~\cite{bastian2}.

With respect to the results reported here, it is remarkable that all the
witnesses used in this paper [Lemma 1 and Eqs.~(\ref{y5wit}, \ref{c5wit},
\ref{r5wit})] were derived from the theory of PPT mixtures. Any PPT mixture
within the considered subclass (e.g., the cluster-diagonal states) will
therefore fulfill the conditions set by the witnesses [e.g.,
Eqs.~(\ref{condition1}, \ref{condition2})] and must be biseparable. In other
words, we have shown that for the families of graph-diagonal states considered 
here, biseparability is equivalent to being a PPT mixture. 

This leads to the question, whether it is generally true that graph-diagonal
states are biseparable if and only if they are PPT mixtures. If this conjecture
were true, it would solve the problem of characterizing multiparticle
entanglement for a huge class of states with an arbitrary number of qubits.
Moreover, for graph-diagonal states the problem can be solved with linear
programming, which is significantly simpler than semidefinite programming
\cite{bastian2} and which could deal with larger qubit systems. However, there
is evidence that the conjecture is not correct, as we explain in the following.

First, note that when looking for a decomposition of a graph-diagonal state into
biseparable (or PPT) states, one can assume that the terms in the decomposition
are also graph-diagonal. If one finds a decomposition where this is not the
case, one can always apply the local depolarization map explained in Section II
between Eqs.~(\ref{eq5}) and  (\ref{stabs}) . The graph-diagonal state is
invariant, but terms in the decomposition which are not diagonal, become
diagonal after application of the map. Since this operation is local, the state
remains biseparable or PPT.

Therefore, if any graph-diagonal state which is PPT with respect to a given
bipartition, is also separable with respect to the same bipartition, the
conjecture would be correct. However, this is not always the case. Examples can
be given from bound entangled states known in the literature
\cite{benatti,piani}. For instance, consider the four-qubit cluster-diagonal
state
\bea
\tilde{\vr}&=&\frac{1}{6}
(
\ketbrapro{\!+\!+\!--}+
\ketbrapro{\!-\!+\!+-}+
\ketbrapro{\!-\!-\!-+}\phantom{++}
\nonumber
\\
&&
+\ketbrapro{\!+\!-\!+-}+
\ketbrapro{\!+\!-\!--}+
\ketbrapro{\!+\!-\!-+}
).
\label{gegenbeispiel}
\eea
where the qubits are as usually written in the order $ABCD.$ Though this state 
is PPT with respect to the $AD|BC$ partition it is entangled with respect to the 
same partition. This can be seen as follows: In Ref.~\cite{benatti} the 
four-qubit state $\hat{\vr}$ 
\begin{align}
\hat{\vr} &=\frac{1}{6}(
\ketbrapro{\Phi^+}_{AB}\otimes\ketbrapro{\Psi^-}_{DC}+
\ketbrapro{\Psi^+}_{AB}\otimes\ketbrapro{\Psi^+}_{DC}
\nonumber
\\
&
+\ketbrapro{\Psi^-}_{AB}\otimes\ketbrapro{\Phi^-}_{DC}
+\ketbrapro{\Phi^-}_{AB}\otimes\ketbrapro{\Psi^+}_{DC}
\nonumber
\\
&
+\ketbrapro{\Phi^-}_{AB}\otimes\ketbrapro{\Psi^-}_{DC}+
\ketbrapro{\Phi^-}_{AB}\otimes\ketbrapro{\Phi^-}_{DC}
),
\label{gegenbeispiel2}
\end{align}
was investigated, here, $\ket{\Psi^\pm}=(\ket{01}\pm\ket{10})/\sqrt{2}$
and $\ket{\Phi^\pm}=(\ket{00}\pm\ket{11})/\sqrt{2}$ are the Bell states.
It was shown that this state is is PPT, but still entangled with respect to the
$AD|BC$-partition. Since the Bell states can be interpreted as two-qubit graph
states, this is a graph-diagonal state. Adding a connection between the qubits
$B$ and $C$ via a controlled phase gate leads to the four-qubit cluster-diagonal 
state $\tilde\vr$ in Eq.~(\ref{gegenbeispiel}) which is PPT for
the $AD|BC$-partition, but nevertheless entangled. Similar examples could be
constructed for higher numbers of qubits \cite{piani}. This demonstrates that
for higher numbers of qubits there might be graph-diagonal states which are PPT
mixtures, but nevertheless genuine multiparticle entangled.

\section{PPT mixtures and the five-qubit $Y_5$ state}

In the previous section, we have argued that one cannot, in general, expect that
the criterion of PPT mixtures is necessary and sufficient for entanglement in
graph-diagonal states. In this section, however, we show that for graph-diagonal
states associated to the five-qubit $Y_5$ graph, the criterion of PPT mixtures
is, in fact, necessary and sufficient for entanglement.

The basic idea of our proof is that for the $Y_5$ state, bound entangled states
such as those given in Eqs.~(\ref{gegenbeispiel}, \ref{gegenbeispiel2}) play no
role in the decomposition. To start, note that the state $\tilde\vr$ in
Eq.~(\ref{gegenbeispiel}), despite being entangled for $AD|BC$, is biseparable
and a decomposition can directly be written down with the help of Lemma 2. This
highlights an interesting detail in the proof of Theorem 3: for the biseparable
decompositions identified in Lemma~2, only the bipartitions $A|BCD$ (and
permutations) and $AB|CD$ have been used, but not the bipartitions $AD|BC$ and
$AC|BD.$ Interestingly, there is a fundamental difference between these types of
bipartitions. For the first set, the entanglement between the two partitions in
the pure graph state $\ket{C_4}$ is equal to one Bell-pair (or one e-bit). This
can be seen from the Schmidt decomposition of $\ket{C_4}$ with respect to that
partition (where the Schmidt coefficients are both $1/\sqrt{2}$). Alternatively,
this follows from the structure of the graph (since, after suitable
transformations which are local for the given bipartition there is only one
connection between the parties). In the second set (the partitions $AD|BC$ and
$AC|BD$) the entanglement between the partitions is equal to two Bell pairs.
Consequently, we refer to the first type of bipartitions as 1BP and the second
type as 2BP.

We can now formulate a fundamental observation linking PPT to separability. If
we have a graph-diagonal state and a 1BP partition, then the PPT criterion is
clearly necessary and sufficient for separability since, after suitable local
operations, the state can be viewed as a two-qubit state
\footnote{To give a precise argument, consider a three-qubit graph-diagonal
state $\vr$ using the linear graph $1$---$2$---$3$  which is PPT with respect to
the $A|BC$-partition. After a controlled phase gate between qubits 2,3 (which is
a local operation for the $A|BC$-partition) the state is transformed to
$\tilde{\vr}=\vr^+_{AB}\otimes \ketbra{+}_{C} + \vr^-_{AB}\otimes
\ketbra{-}_{C}$ where $\ket{\pm} = (\ket{0}\pm \ket{1})/\sqrt{2}$ and the
$\vr^\pm_{AB}$ are two-qubit graph-diagonal states for the graph $1$---$2$.
Since one can deterministically prepare  $\vr^+_{AB}$ and $\vr^-_{AB}$ by
measuring $X$ on the third qubit, both the  $\vr^\pm_{AB}$ must also be PPT and
hence separable. This demonstrates that the original state $\vr$ was also
separable. A similar argument is used in the proof of Lemma 9.}. On the other
hand, 
for a 2BP partition, this is definitely not the case, as the examples in
Eqs.~(\ref{gegenbeispiel}, \ref{gegenbeispiel2}) demonstrate. We formulate this
as follows:

{\bf Corollary 7.} For any biseparable cluster-diagonal state of four qubits
there is a decomposition using 1BP partitions only. Consequently, when looking
for a decomposition for a given {four-qubit} cluster-diagonal state, it suffices
to consider 1BP partitions only.

This statement directly follows from the proof of Theorem 3, since only 1BP
partitions have been used there. It is straighforward to generalize this
slightly as follows:

{\bf Lemma 8.} Let $\varrho$ be a four-qubit graph-diagonal state for an
arbitrary graph, which is PPT with respect to a given partition. Then, $\vr$ can
be written as a PPT mixture using 1BP partitions only.

{\it Proof.}
First, note that the statement is only non-trivial if the given partition is
2BP. Furthermore, note that up to local complementations (or local unitaries)
there are only two different graphs, the $GHZ_4$ graph and the linear cluster
graph $C_4$. For the $GHZ_4$ graph any bipartition is 1BP. For the cluster
graph, however, being PPT for the given partition implies that the state is
biseparable since then the expectation values of the witnesses in Lemma~1 are
nonnegative. Then the claim follows from Corollary~7.
\qed

In order to apply similar ideas to the $Y_5$ state we need {to generalize the
above statement} to five qubits. For five qubits, one can similarly consider 1BP
and 2BP partitions. There are no partitions with three Bell pairs, as this would
require at least six qubits.

{\bf Lemma 9.} Consider a connected five-qubit graph and a two- vs.~three-qubit
partition where one of the qubits in the three-qubit part of the partition is
connected with only one other qubit in the same three-qubit part. Let $\vr$ be a
graph-diagonal five-qubit state being PPT for the given partition. Then, $\vr$
is a PPT mixture using 1BP partitions only.

First, to give an example where the condition on the graph holds, consider the
$Y_5$ graph in Fig.~1(d) and the $ACE|BD$ (or $135|24$) partition. Then, the
qubit E (or 5) is connected only with one qubit in the same part of the
partition, namely the qubit C (or 3). Thus, the condition on the graph is
fulfilled. Note that in this case the partition is a 2BP partition, so the
statement of the Lemma is not trivial.

{\it Proof.}
To prove Lemma 9, we assume {without loss of generality} that the partition is
the $AB|CDE$ bipartition and $E$ is the singular qubit connected only with qubit
$D$. By a suitable local transformation (acting on $DE$ only), one can decouple
the qubit E from the rest. This means that the state $\vr$ is transformed to
$\hat\vr = \vr^+ \otimes \ketbra{+}_E + \vr^- \otimes \ketbra{-}_E$ where the
$\vr^\pm$ are unnormalized states on the qubits $ABCD.$ Since $\hat\vr$ is PPT
with respect to the $AB|CDE$ partition, the states $\vr^\pm$ must also be PPT
with respect to the $AB|CD$ partition.  Otherwise, it would be possible to
generate non-positive partial transpose (NPT) entanglement from a PPT state by
measuring $E$ and distinguishing between $\vr^+$ and $\vr^-$. This is known to
be impossible \cite{oldhoro}. Hence, according to Lemma 8, the states $\vr^+$
and $\vr^-$ form PPT mixtures with respect to 1BP partitions on the qubits
$ABCD.$ Reconnecting the qubit $E$ on the side of $D$ in a 1BP partition on 
$ABCD$ leads to a bipartition on five qubits, which is 1BP, even if $E$ is again
connected with $D$. This immediately induces a PPT mixture of $\vr,$ where only
1BP partitions occur in the decomposition.
\qed

We can now formulate our main result for the $Y_5$ state where all two- vs.
three-qubit bipartitions (2-3-partitions) are either 1BP or fulfill the
conditions of Lemma 9.

{\bf Theorem 10.} 
A $Y_5$-graph-diagonal state is biseparable, if and only if it is a PPT mixture.

{\it Proof.}
Clearly, a biseparable state is also a PPT mixture, which proves one direction
of the claim. Concerning the other direction, let us consider a PPT mixture and
recall that if a PPT mixture is graph-diagonal, then the terms in the mixture
can be chosen to be graph-diagonal as well \cite{bastian2}. We will argue that
the terms belonging to the 2BP partitions in the PPT mixture of the state $Y_5$
can be written as mixtures of 1BP partitions. For this we will make use of
Lemma~9. 

The only candidates for 2BP partitions are the 2-3-partitions, as the
1-4-partitions are automatically 1BP. For the $Y_5$ graph several 2-3-partitions
are in 2BP, however, all fulfill the conditions of Lemma 9: the $ACE|BD$
partition has already been discussed and the $AE|BCD$ partition satisfies the
condition directly. The $BC|ADE$ partition is 2BP and does not fulfill the
condition directly. Nevertheless, after a local complementation on qubit $C$ and
then on qubit $E$, the qubit $D$ is left connected only with qubit $E$ so that
it meets the conditions of Lemma 9. The same sequence of local complementations
can be applied to the $AC|BDE$ partition to show that it too meets the
conditions of Lemma 9. These are, up to symmetries, all of the 2BP partitions.
This implies that all 2BP partitions of $Y_5$ {meet the conditions of Lemma 9}
and thus the PPT criterion is necessary and sufficient to demonstrate
multiparticle entanglement proving the claim. 
\qed

From the proof of Theorem 10 it also follows that the search for the
decomposition into PPT states can be restricted to 1BP partitions. In practice,
one can easily modify the existing algorithms \cite{pptmixture} to consider 1BP
partitions only, which would even make the numerical program simpler. 

An extension of this theorem to other five-qubit graphs is not straightforward.
For instance, for the linear cluster graph [Fig.~1(e)] the partition $BD|ACE$ is
2BP but does not fulfil the conditions of Lemma 9 even after local
complementation. However, this bipartition appears relevant in the decomposition
since it is used in $\eta_3$ of Eq.~(\ref{lc5discussion}).

\section{Generalizations to more than five particles}

So far, we have investigated the separability problem for graph-diagonal states
with up to five qubits and found solutions for many important cases. In this
section we provide two examples that demonstrate how our results can be used to
investigate entanglement in graph-diagonal states with even larger number of
qubits.

\subsection{A generalization of Theorem 10}

In our first example we consider the $Y_N$-state, a generalization of the
$Y_5$-state, shown in Fig.~\ref{fig:thresh2}(a). For this family of states we
can generalize Theorem 10 and show that the criterion of PPT mixtures is
necessary and sufficient. We need the following lemma:

\begin{figure}[t]
\centering
\includegraphics[width=\columnwidth]{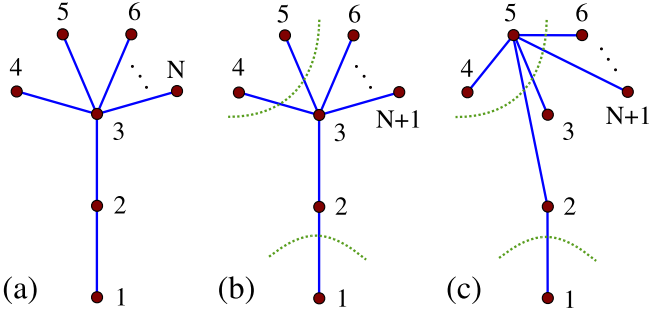}
\caption{(a) The graph of a $Y_N$-state, a possible generalization of the
$Y_5$-state. (b) A possible 2BP partition, here we have chosen $i=5$ and $j=4$.
(c) After local complementation on the qubits $3$ and $i=5$, the qubit $5$ is
the ``central'' qubit. The qubits $4$ and $5$ are then in the condition required
to fulfill Lemma 9. See text for further discussion. }
\label{fig:thresh2}
\end{figure}

{\bf Lemma 11.} Let $\vr$ be a $Y_N$-graph-diagonal state with $N\geq~5$ 
and consider a 2BP partition. Then, if $\vr$ is PPT with respect to that
partition, it can be written as a PPT mixture using 1BP partitions only.

{\it Proof.} We prove the statement by induction. The base case for the
induction, $N=5$, has 
already been proven. For the inductive step, consider the $Y_{N+1}$ graph and a
2BP partition [see Fig.~\ref{fig:thresh2}(b)]. We denote the two parts of the
bipartition as $M$ and $\overline{M}$. One can directly see that qubits $1$ and
$2$ must belong to different parts of the partition, otherwise the partition is
only 1BP. We assume that $1 \in M$ and $2\in\overline{M}.$ Furthermore, of the
remaining qubits $\{3, ... ,N+1\}$ there must be at least one belonging to $M$
and at least one belonging to $\overline{M}.$ Otherwise, the partition would be
a one vs.~$N$ partition, which can never be a 2BP partition. Since $N \geq 5$
there must be either two qubits from the set $\{3, ... ,N+1\}$ in $M$ or two
qubits from the set $\{3, ... ,N+1\}$ in $\overline{M}.$ Let us assume that the
two qubits $i,j \in \{3, ... ,N+1\}$ belong to $M$.

Now we apply local complementation on qubit $3$ and then on qubit $i$ [see
Fig.~\ref{fig:thresh2}(c)]. If $i=3$ this changes nothing. Otherwise, the graph
is transformed so that qubits $3$ and $i$ are interchanged, hence the qubit $i$
is afterwards the ``central'' qubit. Qubit $j$ is now only connected to the
qubit $i$, and both qubits belong to the same part of the partition.

This, however, is exactly the situation as described in Lemma 9. As in the proof
of Lemma 9 we can decouple the qubits $i$ and $j$, and the remaining two states
$\vr^\pm$ are $Y_N$-graph-diagonal states of $N$ qubits, which are PPT with
respect to the given 2BP partition. By the induction hypothesis, these states
are PPT mixtures with respect to 1BP partitions. Translating this backwards by
inserting again the previously deleted connection finally proves the claim.
\qed

Having proven Lemma 11 we can formulate: 

{\bf Theorem 12.} 
A $Y_N$-graph-diagonal state is biseparable, if and only if it is a PPT mixture.

{\it Proof.} The proof is essentially the same as that of Theorem~10.
We only have to consider 1BP partitions according to Lemma~11 and for them
PPT is necessary and sufficient. Note that for the $Y_N$ graph there are only
1BP and 2BP partitions, 3BP partitions are not possible.
\qed

As with the case of the $Y_5$ state discussed after Theorem 10, one can simplify
the search for PPT mixtures in the $Y_N$ state by concentrating only on the 1BP
partitions. This makes it possible to determine separability for
$Y_N$-graph-diagonal states for larger values of $N$ though the number of 1BP
partitions still grows fast.

\subsection{Biseparable decompositions for linear cluster states}
For our second example of separability conditions for graph states of more than
five particles, let us discuss the six-qubit linear cluster state mixed with
white noise. Our goal is to show that the criteria used in this paper allow
estimates of separable regions in a simple way even for graph-diagonal states
with many qubits, and the resulting estimate is quite accurate.

First, in a straightforward  generalization of Lemma 2, many pairs of the form
$\sigma=\ketbrapro{\!+\!+\!+\!+\!++}+\ketbrapro{ijklmn}$ are separable, the
exceptions are the 44 states $\chi_i=\ketbrapro{\!+\!+\!+\!+\!+ +} +\eta_i,$
with $\eta_1=\ketbrapro{\!-\!jk\!-\!++}$, $\eta_2=\ketbrapro{\!-\!jkl\!-+}$,
$\eta_3=\ketbrapro{\!-\!jklm-}$, $\eta_4=\ketbrapro{\!+\!+\!-jk-}$,
$\eta_5=\ketbrapro{\!+\!-jkl-}$, or $\eta_6=\ketbrapro{\!+\!-jk\!-+}$.
Furthermore, using the fact that the state from Eq.~(\ref{remainingstate}) is
separable, one can directly find a biseparable decomposition of
\be
\vr(p)=p\ketbra{C_6} + (1-p)\frac{\eins}{64}
\ee
for $p=11/43 \approx 0.256.$ Since the state $\vr(p)$ is known to be entangled
for $p>51/179 \approx 0.285$ \cite{bastian2} the real threshold cannot be much
higher and this simple estimate already delivers a good approximation.

This method of constructing biseparable decompositions in the graph basis of
linear cluster states can be generalized to an arbitrary number of qubits.

\section{Conclusion}

In conclusion, we have considered the problem of detecting genuine multiparticle
entanglement in graph-diagonal states for four and five qubits and we have
provided complete solutions for many important cases. In addition, we showed how
these results allow us to gain insight into this problem for larger numbers of
qubits. Since our results deliver optimal criteria, they can be used to test the
strength of other entanglement criteria.

We believe that the study of entanglement in graph-diagonal states is an
interesting and fruitful area of research. These states are extremely important
from the point of view of experiments in quantum information science and, from
the theoretical point of view, these states have an elegant description in the
stabilizer formalism. Future work would involve formulating detection criteria
for other entanglement-related problems for this class of states. For instance,
are there necessary and sufficient criteria for other forms of entanglement?
This may include the notion of full separability (for first results see
Ref.~\cite{graphdiagonal2}) or the task of entanglement distillation. Can some
multiparticle entanglement measures be computed for these types of mixed states?
Both of these questions, and many others like them, are in need of further
research.

We thank M.~Ali, M.~Hofmann, M.~Kleinmann and S.~Niekamp 
for discussions. This work has been supported by
the Austrian Science Fund (FWF): Y376-N16 (START prize), 
the EU (Marie Curie CIG 293993/ENFOQI), and 
the MITRE Innovation Program, Grant 07MSR205.

\end{document}